# Why engineers are right to avoid the quantum reality offered by the orthodox theory

Xavier Oriols, *Member, IEEE*, David K. Ferry, *Life Fellow, IEEE*

*Abstract*— A proper knowledge of the reality of a physical theory is important to get an understanding of empirical phenomena. Despite its extraordinary predictive successes, the orthodox (also known as the Copenhagen) theory provides an indigestible definition of quantum reality: objects (such as electrons) become part of a nebulous many-particle wave function with no properties at all, unless the property is explicitly measured. To make matters worse, orthodox theory does not define measurements in a clear way. This view of reality is foreign to most modern engineers who assume that quantum objects, like classical objects, always have real properties, independent of any measurement. Despite being contrary to the orthodox theory, the intuition of engineers is not in conflict with other quantum theories, where the observer plays no fundamental role. Good quantum intuition needs to be based on a correct knowledge of the fundamental elements of the quantum theory that is being used. We argue that engineers are actually led to the *natural* quantum reality offered by these alternative approaches.

## I. INTRODUCTION

WE are currently in the middle of a second quantum revolution where the rules discovered a century ago to understand the quantum world are being applied to develop new quantum technologies [1]. Yet, most of the understanding of quantum physics is developed from the orthodox (also known as Copenhagen) interpretation of the quantum phenomena [2], [3]. However, the orthodox theory has important difficulties in providing an intuitive view of the quantum technologies because it states that a quantum object only has real properties when it is measured by an observer. This view of the reality of quantum objects is unnatural to engineers, and thus is ignored by them when analyzing their real-world devices. What is *natural* to engineers is to imagine that the reality of an object (their properties) is independent of whether the object is measured or not.

There are other quantum theories [4]-[10] that have no empirical contradiction with the results of the orthodox theory, yet are fully compatible with the engineer's *natural* view of reality. Engineers, consciously or unconsciously, avoid the orthodox reality and adopt the much more familiar notion of reality, independent of the observer, offered by these other quantum theories [4]-[10]. Does the adoption of this *natural* reality for quantum objects create any contradictions in quantum physics? Not at all. Here, we will examine the reasons why engineers avoid the orthodox notion of reality and we will justify the scientific viability of using a more *natural* view of quantum phenomena, where the observers play no fundamental role. To the best of our knowledge, the ideas explained in this paper are not common in the literature because the "*shut up and calculate*" attitude [11] has been impressed upon young engineers.

First, however, we give a brief summary of the essence of the orthodox theory and its dependence upon the observer. Then, we show how other approaches (or interpretations) support the engineer's choice for designing quantum technologies, and then turn to some examples of this in quantum electron devices.

### A. Reality from orthodox quantum mechanics

In the orthodox theory, an isolated quantum object (like an electron) is described by a wave function $\psi(\vec{r}, t)$, which is the essential (ontological) element of the orthodox theory [3]. In the non-relativistic regime in which we are interested[1], the evolution in time of such wave functions is given by the Schrödinger equation [12]. In the orthodox theory, the reality of the property A of an electron (such as its position or its velocity) is defined through the eigenvalue-eigenstate link [2], [3]. Each property A of the electron is defined through an operator $\hat{A}$. Such an operator has different eigenfunctions $\psi_a(\vec{r})$, each with its own eigenvalue, which is the value of the property A of the electron when described by that specific eigenfunction. Since the Schrödinger equation is linear, in general, it is possible for the isolated system to be described by a sum (linear superposition [3]) of the different possible eigenfunctions.

This work was supported in part acknowledges funding from Fondo Europeo de Desarrollo Regional (FEDER), the QUANTUMCAT project 001-P-001644, the 'Ministerio de Ciencia e Innovación' through the Spanish Projects No. RIT2018-097876-B-C21, the European Union's Horizon 2020 research and innovation program under Grant Agreement No. Graphene Core3 under grant 881603 and under the Marie Skłodowska-Curie Grant Agreement No. 765426 (TeraApps).
Xavier Oriols is with the Departament d'Enginyeria Electrònica, Universitat Autònoma de Barcelona 08193-Bellaterra (Barcelona), Spain (e-mail: xavier.oriols@uab.es).
David K. Ferry is with the School of Electrical, Computer, and Energy Engineering, Arizona State University, Tempe, AZ 85287 USA (e-mail: ferry@asu.edu).

---

[1] Throughout this paper, we will focus only on non-relativistic quantum mechanics. We are not discussing in a universal theory of everything. We just want to find an appropriate theory to explain the performances of quantum electron devices. See the conclusion section for an enlarged discussion.

For example, let us assume that we want to measure the location of an electron, to the left or the right of a barrier, during a tunnelling process. At the end of the tunneling process and before the measurement, the whole wave function contains the eigenfunction $\psi_L(\vec{r},t)$ for a particle to the left of the barrier (reflected component), and the eigenfunction $\psi_R(\vec{r},t)$ of the particle to the right of the barrier (transmitted component). However, we know from measurements in the laboratory that the charge of an electron ($q = -1.6 \cdot 10^{-19}$ Coulombs) cannot be located at both sides of the barrier. The electron is either transmitted or reflected, but not half-transmitted, and half-reflected! A measurement of the location will define on which side the electron exists.

To understand why, in a measurement process, we can obtain a well-defined location of the charge of an electron, either in the left (emitter or source) contact or in the right (collector or drain) contact, the orthodox theory needs to introduce an extra equation (different from the Schrödinger equation) named the collapse law to describe the state reduction [2], [3]. This new equation of motion can only be invoked during a measurement process and it selects only a single eigenfunction from the superposition of transmitted and reflected components[2]. For example, if the measured location of the charge is on the left, the evolution of the wave function during the measurement is $\psi(\vec{r},t) \rightarrow \psi_L(\vec{r},t)$. Then, the property of the electron (in our example, its left or right location) becomes well-defined. If we stop measuring the system, the new (post collapse) wave function evolves following the Schrödinger equation into a new superposition $\psi_L(\vec{r},t) \rightarrow \psi(\vec{r},t)$. The previous well-defined location of the electron is lost; it is undefined once again.

So, in general, the electron has no well-defined position and no well-defined velocity (except if the wave function is an eigenfunction of one of these properties). Does the electron exist when it has none of its properties well-defined? If we use the *natural* definition of existence inherited from our common sense (particles exist because they have some well-defined properties; in particular, the position in our ordinary 3D space), the orthodox theory implies an intermittent reality for electrons. In fact, if we are dealing with a systems with many electrons, each one with the position degrees of freedom indicated by $\vec{r}^i$, then most of the time the electron is just a part of the nebulous many-particle wave function $\Psi(\vec{r}^1, \vec{r}^2, \ldots, \vec{r}^N, t)$ *living* in a huge 3N dimensional configuration space.

Let us emphasize that the orthodox theory is not stating that we ignore the properties of the electrons because we have experimental limitations to access information about their location. No. *The orthodox theory is directly stating that the properties themselves are undefined, unless the property is measured.* This dependence of the quantum reality on the fact of measurement was brought to quantum mechanics by Bohr himself [13], and presumably follows a philosophical basis popular in the late nineteenth century (and even into post world war I Europe) inspired by what was called the Vienna Circle. It is famously represented by Ernst Mach [14], who didn't believe in atoms because he couldn't see them. The orthodox theory has persisted in negating the reality of what cannot be explicitly measured to this day without much public controversy because, as Gell-Mann put it [15]: "*The fact that an adequate philosophical interpretation [of quantum mechanics] has been delayed so long is no doubt caused by the fact that Niels Bohr brain-washed a whole generation of theorists into thinking that the job was done 50 years ago.*"

### B. Reality in Alternative Views of Quantum Mechanics

As indicated in the introduction, there are other quantum theories which do not rely upon the observer for defining the reality. These alternatives theories appeared from the very beginning of quantum mechanics with the works of de Broglie [4], Madelung [5] and Kennard [6] who pointed out that quantum phenomena, although more exotic than classical ones, does not require one to throw away our classical notion of reality. As Kennard put it [6]: "*...each element of the probability moves in the Cartesian space of each particle as that particle would move according to Newton's laws under the physical force plus a 'quantum force'....*" The Bohm theory[3] [9], [16], [17] is one of these alternatives quantum approaches. We will adopt it as an example on how the dynamics of quantum objects can be fully understood without negating their reality.

According to Bohmian mechanics, an electron has a particle and a wave nature. The particle-like nature is given by a well-defined (property) position in our ordinary 3D space, $\vec{r}(t)$, at any time $t$. The wave-like nature is given by the same wave function $\psi(\vec{r},t)$ discussed previously, which guides the particle via a velocity field defined from the classical force and the quantum force, where the latter is derived from the squared magnitude of the wave function. Hence, the particle will have a well-defined trajectory.

The initial position $\vec{r}(t_0)$ has to be fixed according to the probability distribution $|\psi(\vec{r},t_0)|^2$. As happens when we prepare several identical experiments with a gas of classical particles (that we can fix the volume and the temperature, but not the initial positions of the particles), for identically prepared quantum experiments with exactly the same wave function $\psi(\vec{r},t)$, we cannot fix the initial positions of the particles. Then, when repeating exactly the same quantum experiment (with the same wave function) we can have different trajectories that involve different outputs. This is the origin of the quantum uncertainty according to the Bohm theory [16]-[18].

The measurement of the property A of the electron requires an interaction between the electron itself and the measuring apparatus. As such, new degrees of freedom have to be included in the many-particle wave function and new trajectories considered. Thus, there is a back action of the apparatus on the electron, and vice versa. In technical words, the Bohmian theory is contextual [16], [17], [19]. As happens with the orthodox theory, the wave function that guides now the electron is the many-particle wave function $\Psi(\vec{r}^1, \vec{r}^2, \ldots, \vec{r}^N, t)$ in the 3N dimensional space. However, the crucial point is that the electron of the system has well-defined properties in our ordinary 3D space before and after the measurement. Even though the measurement can modify the electron trajectory, it has no special role in defining the reality of the electron. In this sense, this theory is named a "*quantum theory without observer*" and its view of the quantum reality is very close to the classical reality and, thus, very *natural* for us. As we will see in next section, this is exactly how engineers think on quantum electron devices.

---

[2] Here, for simplicity, we are assuming a strong or projective measurement. Later, in section II when discussing a real electron device, we will show that this tunneling process is, in fact, a weak or indirect measurement.

[3] The Bohm theory is also known as in the literature as the Bohmian mechanics, the de Broglie-Bohm theory, the pilot-wave model, and the causal interpretation of quantum mechanics.

## II. HOW ENGINEERS UNDERSTAND QUANTUM ELECTRON DEVICES?

We argue in this section that engineers, in fact, ignore the orthodox version of the quantum reality when explaining their work. Engineers have developed their own way of thinking of quantum phenomena, where the observers play no role at all. The (very) bad news for the orthodox theory is that this engineering way of thinking is not a naïve one, but it is able to satisfactorily and accurately explain all quantum phenomena. In fact, they are following the alternative view of quantum mechanics seen above. Let us explain this conclusion by considering a real semiconductor device as our quantum system, and the electrical current at frequencies close to one THz as the property A of the quantum system that we are looking for.

### A. A natural explanation from a Bohm-like reality

In Fig. 1(a), we illustrate a typical graphene field-effect transistor (GFET) which is a research-level electron device where transport needs to be understood through quantum laws because of the possible presence of spatial quantization and the appearance of Klein tunnelling [20]. The passage of electrons from the source and drain contacts, through the 2D graphene structure, is controlled by the voltage in the gate contacts, just as in any other FET.

In Fig. 1(b), we see three different current-voltage characteristics of the GFET. The conduction band (CB) corresponds to electrons above the Dirac point and the valence band (VB) below (the Dirac point is the point of zero momentum where the CB and VB touch in graphene, which has no band gap). The blue (square) dashed curve in Fig. 1(b) corresponds to the scenario where electrons are injected from both conduction band (CB) and valence band (VB) from the zero-gap graphene [21]. Contrary to normal transistors, we see in the blue curve that there is no saturation current, since applying more voltage between source and drain leads to more carriers travelling from source to drain, from the VB in the source contact to the CB in the drain contact [22]. The electrons in the VB at the source contact will be able to reach the CB in the drain contact because they can be transmitted through any potential shape with probability close to one due to Klein tunnelling. On the other hand, in the dashed light blue (diamond) curve, we allow only injection from the CB. Then, current saturates because after the voltage reaches the Fermi energy value, the same fixed number of electrons from the conduction band are injected independently of the applied voltage. See the bottom inset. Finally, in the orange solid (up triangle) curve, dissipation due to acoustic and optical phonons are considered. This last curve confirms that the electron transport is mainly ballistic [21].

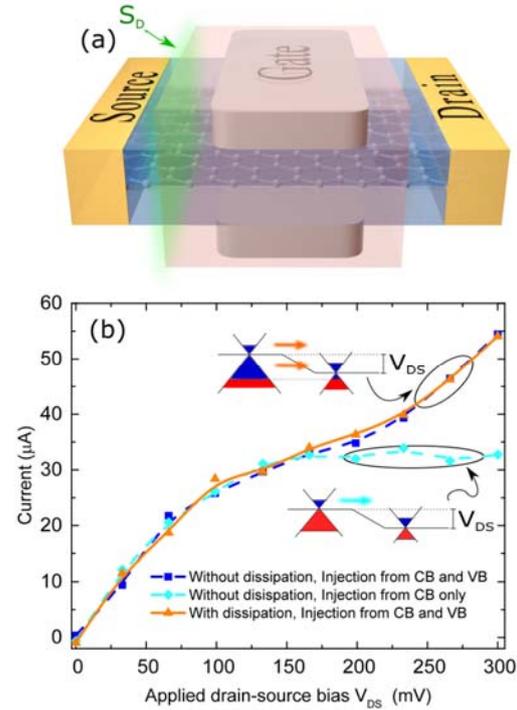

Fig. 1. (a) Schematic view of a dual-gate graphene field-effect transistor (GFET). The central (pink) region corresponds to the device active region of the GFET and the surface $S_D$ where the electrical current is evaluated. (b) Simulated current-voltage characteristic for three simulated GFETs with bottom and top gates biases at 0 V. The dashed lines are for the ballistic transport with the dark blue (square) one represents normal graphene injection (electrons injected from both the CB and VB) current-voltage characteristic and the light blue (diamond) line represents only electrons from the CB are injected. In the orange solid (up triangle) curve, dissipation due to acoustic and optical phonons are taken into account. The insets sketch different energy profiles for different scenarios.

In next subsection, we analyse if such typical GET explanation can be supported by the orthodox reality. The answer will be negative as can be anticipated by the fact that we have been able to explain the performance of the GFET as if the electrons were real (inside the active region) without any mention of their measurement. But, as we see in section IA, the orthodox version of quantum mechanics explained in textbooks links the reality of electrons to their explicit measurement.

### B. The difficulties of the orthodox-like reality

Let us start by pointing out that the simple (and innocent) sentence, "*carriers traveling from the source to the drain*", is forbidden in orthodox theory. Without the direct measurement of electron positions, the orthodox theory says that the electron has no location, no velocity at all. An obedient orthodox follower has to explain the GFET performance using the tools offered by direct observation (measurement). If we take the orthodox theory seriously, we have to accept that the plot in Fig. 1(a), and all such plots found in all textbooks, is unacceptable as we have well-defined locations of the differents parts/atoms of the nanoscale electron device without explicitly measuring them.

Since the orthodox reality depends on the measurement, and the measurement is modeled by an operator, the engineer has to establish the operator that corresponds to the measurement of the current in the GFET. At first sight, it seems that we can model the orthodox measurement by collapsing the wave function of the electrons in the active region into an eigenstate of the current operator, as dictated by the orthodox eigenstate-eigenvalue link.

When blindly following this recipe, a curve like in Fig. 2(b) requires a continuous measurement. If we make a quantum measurement immediately after a first one, we get the same eigenvalue (the same value of the current), not the fluctuations seen in the figure. This contradiction has it origin in the so-called quantum Zeno paradox [23].

Another problem that appears when trying to model the measurement of the current by an operator $\hat{A}$ is that the current plotted in Fig. 1(b) is the DC current which is directly related to the particle current, defined by the number of electrons crossing from left to right minus the electron traversing the device in the opposite direction per unit time. However, this particle current is not what is measured in the ammeter of Fig. 2(a). The measured current corresponds to the plot of Fig. 2(b) which includes the particle and displacement current. The displacement current is proportional to the time-derivative of the electrical field [24], [25]. It is not obvious how to define eigenfunctions for the displacement current.

The reason why we are looking for the proper orthodox operator $\hat{A}$ is because we want to recover the reality of electrons crossing the active region by means of the measurement of the total electrical current. The drama, as we explain below, is that a proper orthodox measurement of the total current implies negating the reality of the electrons in the active region at any time. As shown in Fig. 2(a), the measurement of the electrical current takes place in the ammeter located far from the active region of the device [24], [25]. There is no measuring apparatus on the surface $S_D$ of the GFET active region in Fig. 1(a). The ammeter, by construction, effectively measures the electrical current flowing through the suface $S_A$ in Fig. 2(a). Then, the orthodox collapse only affects the degree of freedom related to the ammeter [26]. Due to the interaction among electrons in the ammeter, cable and active region, the output current given by the ammeter provides indirect (but valuable) information of the electrical current of our electrons inside the active region. This type of indirect measurement of the active region is named weak measurement in the literature [26]. It is different from the required direct or strong measurement of the electrons inside the active region, which is the mandatory mesasuremtn to ensure that they have well-defined orthodox properties (eigenvalues). In weak measurements, the electrons in the active region suffer a small perturbation due to small backaction from the collapsed electrons in the ammeter. In fact, the strong (and bizarre) collapse mentioned in section I.A for the electrons traversing the active region, $\psi(x,t) \to \psi_L(x,t)$, does not occur in a weak measurement. This means that the many-particle wave function of the electrons inside the GFET after the weak measurement is still a superposition of eigenfunctions. At first sight, avoiding the collapse of the wave function in the active region seems a nice scenario to recover the faith in the orthodox view of the quantum reality. But, the electrons in the active region are always a superposition of eigenfunctions without well-defined properties. As Ernst Mach [14] negated the reality of the atoms because he couldn't see them, the orthodox theory negates the reality of the electrons inside the active region because they are not (*strongly*) measured. How can we develop an orthodox intuition about the dynamics of electrons inside the active device region, if the electrons are not located there? Orthodox quantum mechanics does not describe electrons inside the active region as matter moving in 3D space (as Bohm trajectories did), but as a nebulous many-particle wave function in the 3N configuration space (with simultaneous reflected and transmitted components for all of them).

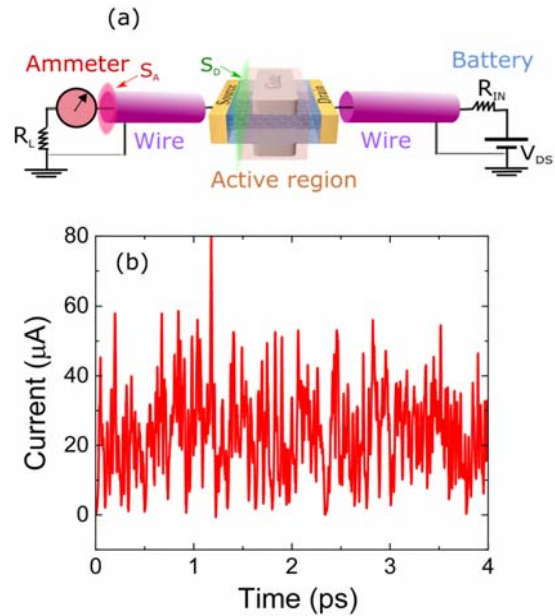

Fig. 2. (a) Schematic plot of the active region of the GFET plus the typical set-up for measuring the electrical current with the surfaces $S_D$ where the electrical current is evaluated and $S_A$ where the electrical current is effectively measured. (b) Simulated total (particle plus displacement) electrical current as a function of time for the GFET of Fig 1 when the bottom and top gates are biased at 0 V and the drain bias is 0.06 V. The type of simulation is without dissipation and from the CB and VB. It corresponds to the dark blue (square) dashed line of Fig. 1.

### C. Mixing realities

We have shown that the "*shut up and calculate*" attitude [11] in front of quantum devices (or quantum technologies in general [27]), without worrying about the reality behind the orthodox mathematics, implies difficulties in creating a healthy intuition about them. Here, we show that this attitude can also lead to erroneous computations.

An electron device is a many-particle open system working far from thermodynamic equilibrium, yet still following quantum laws. In this sense, it is one of the most difficult physical systems that humans can try to simulate and forces us to make important approximations (simplifications). One typical strategy for simplifying the computational burden is what is known as multiscale modeling. Some parts are modeled with standard quantum tools, for example the band structure, while others, for example the transport of electrons, are modeled with semi-classical tools [28]. However, one is usually not aware that such multi-scale modeling also involves matching multi-scale-realities of a unique world. How can an orthodox quantum reality of our world be matched with a classical reality? Perhaps, the most common mistake due to this multiscale mismatch of reality appears when trying to combine quantum electron transport with classical Maxwell's equations. Such a self-consistent combination appears necessary, for example, when modelling nano-electronics devices at THz frequencies. In the classical case, the reality of the charge and current is independent of the measurement process while, in the orthodox theory, the charge only becomes real when it is measured. The typical wrong solution to avoid the nightmare of including the quantum measurement in the orthodox quantum modeling is just assuming that the wave function at any time is equal to the electron charge density. However, we have already clarified before that the wave function in a tunneling process just describes left and right probabilities in different experiments.

This multiscale problem is basically a special case of a more general problem known as the quantum-to-classical transition, i.e., the question of how effectively classical systems and well-defined properties (position, energy) for the objects around us emerge from the underlying quantum domain [29]. Most of these practical and conceptual problems go away if we adopt the reality of an observer-less quantum theory and its ontology. Indeed, the use of the observer-less approach described in Sec. IB has already penetrated into practice. The motion of the particles in the GFET above is guided by an additional quantum force, which has already appeared in device modeling [30], [31] and is included in many commercial packages for such simulation. A more advanced incorporation of quantum effects in a MOSFET utilizes a *quantum* potential as an integral correction to the self-consistent potential in the device [32]. Again, this approach is finding its way into commercial software packages for device simulation. In addition, the numerical simulations of this work have been done from the BITLLES simulator using a description of electrons as Bohmian trajectories guided by the wave function [33].

### III. SUMMARY AND FINAL REMARKS

A theory allows one to predict some range of physical phenomena. Such predictions are formally obtained from the outputs of the equations of motion of the essential elements of the theory. Such essential elements make up the reality (the ontology) of the theory. A proper knowledge of the reality behind a physical theory allows us to develop an intuitive understanding of the empirical phenomena described by such theory without explicitly solving the complex mathematical equations of motion. Despite its extraordinary predictive successes, the orthodox theory of quantum mechanics provides an indigestible view of reality: quantum objects (like electrons) are defined as part of a nebulous many-particle wave function with no well-defined properties, until such properties are explicitly measured. To the contrary, the intuitive understanding of quantum phenomena followed by engineers invokes a reality of quantum properties independently of whether or not they are measured. Such a *natural* notion of reality of quantum objects is supported by alternative quantum theories where the observer plays no role.

We also argue that doing quantum computations with a "*shut up and calculate*" attitude [11], while unconsciously using another theory to develop the intuition that explains the computations is a schizophrenic situation. A correct intuition needs to be based on a proper knowledge of the fundamental (ontologic) elements of the theory. One solution, strongly supported by the present authors, is to recognize that the orthodox view is a *effective* theory, whose justification at the fundamental level, comes from a quantum theory, without observers, that supports a *natural* definition of quantum reality. With this point of view, engineers can take the best of both worlds simultaneously, and avoid having to pay the price of an "*incomprehensible*" quantum reality.

We are not demanding here that the true reality of our world is the one provided by the observer-less theory. All physical theories are wrong when applied outside of their regime of validity. What we defend in this paper is that approaches, such as the Bohm theory, are excellent tools for engineers to help understand the quantum world. They exactly reproduce all quantum phenomena and they allow a very familiar and *natural* view of the quantum reality. In the same way, one says that classical mechanics, although incomplete, is a very useful theory upon which architects can build their houses or aeronautical engineers can develop their airplanes. Similarly, engineers also merit a simple and intuitive understanding to develop quantum technologies. There is no point in explaining the quantum world by invoking complicated explanations of the quantum reality, when more *natural* and familiar views are easily available. The *natural* intuition of engineers about the reality of quantum objects, independent of the measurement, is a healthy intuition. It can be perfectly cultivated to anticipate the behaviour of any quantum device without fear of being betrayed by its simplicity. Demystifying the "*incomprehensible*" quantum world is a mandatory step, if our society really wants to move from quantum science to quantum engineering and achieve the expected goals of the second quantum revolution.

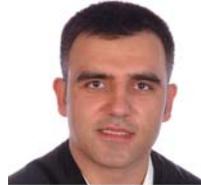

**Xavier Oriols** received his BS and MS in Physics from the Universitat Autónoma de Barcelona (UAB) in 1993 and 1994 respectively. During 1997, he worked at the Institute d' Electronique, Microelectronics and Nanotechnology in Lille (France). He received his doctoral degree in Electronic Engineering from UAB in 1999 with an extraordinary doctoral award. During 2002 he was a visiting professor at the State University of New York (USA). He is full professor of electronics in the UAB and has authored more than 150 contributions to scientific journals and conferences. He is author of the book "Applied Bohmian Mechanics: from nanoscale system to cosmology" and he has developed the device simulator BITLLES (http://europe.uab.es/bitlles). In 2008, he received the prize for young researchers in the framework of the Spanish I3 Program and the prize for research excellence in 2008 and 2010. The research activity of Dr. Oriols combines both his practical interest in quantum nanodevices, and his interest in the foundations of quantum mechanics, in general, and in Bohmian mechanics, in particular. His research covers a wide spectrum, from fundamental issues of physics to practical engineering.

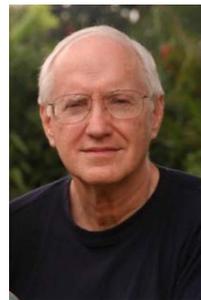

**David Ferry** is Regents' Professor Emeritus in the School of Electrical, Computer, and Energy Engineering at Arizona State University. He was also graduate faculty in the Department of Physics and the Materials Science and Engineering program at ASU, as well as Visiting Professor at Chiba University in Japan. He came to ASU in 1983 following shorter stints at Texas Tech University, the Office of Naval Research, and Colorado State University. In the distant past, he received his doctorate from the University of Texas, Austin, and spent a postdoctoral period at the University of Vienna, Austria. He continues active research. The latter is focused on semiconductors, particularly as they apply to nanotechnology and integrated circuits, as well as quantum effects in devices. In 1999, he received the Cledo Brunetti Award from the Institute of Electrical and Electronics Engineers, and is a Life Fellow of this group as well as Fellow of the American Physical Society and the Institute of Physics (UK). He is the author, co-author, or editor of some 50 books and about 900 refereed scientific contributions.